\documentclass[aps,pra,twocolumn,groupedaddress,showpacs]{revtex4}
\usepackage{graphicx,amsmath,amssymb,natbib}
\usepackage{color}
\usepackage{multirow}
\begin{document}
\bibliographystyle{h-physrev3}

\title{Instability dynamics and breather formation in a horizontally shaken pendulum chain}
\author{Y. Xu$^1$} \author{T. J. Alexander$^{1}$}\email{t.alexander@unsw.edu.au} \author{H. Sidhu$^1$}  \author{P. G. Kevrekidis$^2$} 

\affiliation{$^1$School of Physical, Environmental and Mathematical Sciences, UNSW Canberra, Australia 2610}
\affiliation{$^2$Department of Mathematics and Statistics, University of Massachusetts, Amherst, Massachusetts 01003-4515, USA}

\begin{abstract}
Inspired by the experimental results of Cuevas 
{\em et al.} (Physical Review Letters {\bf 102}, 224101 (2009)), we consider theoretically the behaviour of a chain of planar rigid pendulums suspended in a uniform gravitational field and subjected to a horizontal periodic driving force applied to the pendulum pivots.  We characterize the motion of a single pendulum, finding bistability near the fundamental resonance, and near the period-3 subharmonic resonance.  We examine the development of modulational instability in a driven pendulum chain and find both a critical chain length and critical frequency for the appearance of the instability.  We study the breather solutions and show their connection to the single pendulum dynamics, and extend our analysis to consider multi-frequency breathers connected to the period-3 periodic solution, showing also the possibility of stability in these breather states.  Finally we examine the problem of breather generation and demonstrate a robust scheme for generation of on-site and off-site breathers.
\end{abstract}
 
\maketitle
\section{Introduction}
\label{intro}

Driving of the pivot point of a single pendulum has long been known to lead to parametric driving of the pendulum motion (see e.g. \cite{Nayfeh1979} for a textbook treatment).  The majority of research has focused on vertical driving (see e.g. \cite{BlackburnAJP1992,GrandyAJP1997} and references therein), where the well-known stabilization of the inverted pendulum may be observed \cite{StephensonMPMLPS1908}.  Horizontal driving has received less attention, with early works considering the nature of the periodic solutions \cite{StrubleQJMAM1965} and the appearance of subharmonic excitations \cite{CheshankovJAMM1971}.  More recently, interest in this problem has been revived, with recent results exploring the appearance of chaotic motion \cite{VanDoorenCSF1996}, dynamic stabilization of two off-centre equilibrium points \cite{SchmittND1998} and complex bifurcation behaviour of the period-1 oscillation \cite{JeongJKPS1999}.  

In this work, motivated by the experimental setup of~\cite{CuevasPRL2009},
we consider a setup of pendula subject to a horizontal driving force.
The progression of our study
starts with the case of a single pendulum.  There, we identify  
period-1 solutions near the nonlinear resonance and reveal the familiar foldover effect for a driven pendulum and the appearance of subharmonic, period-3, solutions. This is a building block of relevance towards the study of the 
pendulum chain, to which we  
then turn our attention. In the latter, we find similar results for the period-1 solutions, albeit with the appearance of instability at a critical chain length.  We examine this instability in some detail, identifying the characteristic instability wavenumbers through modulational instability analysis.  We then examine the breather solutions in the system, and in agreement with the results of \cite{CuevasPRL2009} we find families of breather solutions existing in the bistable region.  We examine the instability dynamics and breather formation near this region, and provide a prescription for the generation of breather solutions.  We also explore the possibility of multi-frequency breathers supported by the subharmonic pendulum response and show that these solutions may persist for long times.

Our presentation will be structured as follows. In section II, we
present the relevant model in connection to the experiments 
of~\cite{CuevasPRL2009}, while in section III, we study the single
pendulum dynamical features. Section IV extends this to multi-pendulum
dynamics and the modulational stability, breather existence/stability,
as well as multi-frequency generalizations that can arise.
Finally, section V presents our conclusions, as well as some
potential future challenges.


\section{Model}
\label{model}

\begin{figure}
\includegraphics[bb=0 0 800 600,width=0.9\columnwidth]{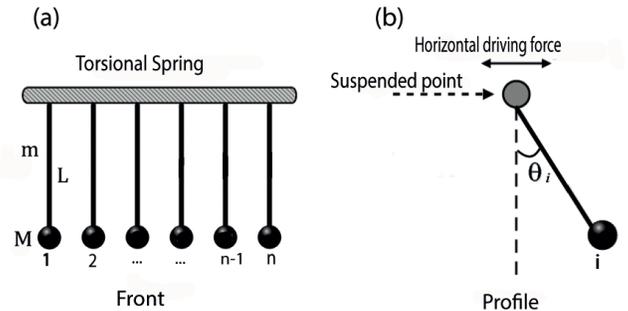}
\caption{Schematic of experimental arrangement for horizontally shaken pendulum chain showing chain in (a) side and (b) profile views.}
\label{fig1}
\end{figure}

The Lagrangian in the absence of damping for a chain of $N$ pendulums subjected to a periodic horizontal displacement of the pivot point with frequency $\omega_d$ and amplitude $A$ (as shown in Fig. \ref{fig1}) takes the form \cite{JoseSaletan}:
\begin{align}
\label{lagrange}
\mathcal{L} &=\sum^{n}_{i=1} \frac{1}{2}(Ml+ m\frac{l}{2})\left[2A\omega_d \sin(\omega_d t)\cos \theta_i \dot{\theta_i} \right] \\ &+ \frac{1}{2}I\dot{\theta_i}^2 + (Ml+m\frac{l}{2}) g\cos\theta_i\nonumber \\
&-\frac{1}{2}\beta\left[(\theta_i-\theta_{i-1})^2+(\theta_i-\theta_{i+1})^2\right] \nonumber
\end{align}
where $m$ is the mass of the thin rod of length $l$ supporting the pendulum bob of mass $M$, and $\theta_i$ and $\dot{\theta}_i$ are the angle relative to vertical and angular speed respectively for the $i$-th pendulum.  The moment of inertia is given by $I = Ml^2 + ml^2/3$, $g$ is the acceleration due to gravity and $\beta$ is the linear coupling between pendulums through a torsion spring.  We consider the case with end pendulums defined by taking $\theta_{N+1} = \theta_N$, $\dot{\theta}_{N+1} = \dot{\theta}_N$ and $\theta_0 = \theta_1$,$\dot{\theta}_0 = \dot{\theta}_1$.  

As discussed in Ref. \cite{CuevasPRL2009} in a physical system there is also on-site damping due to velocity-dependent friction (due to air resistance) and intersite damping due to frictional loss in the torsion spring.  The equations of motion therefore take the form \cite{CuevasPRL2009}:
\begin{align}
\label{physmod}
\ddot{\theta}_n&- \frac{\beta}{I}\left(\theta_{n+1}+\theta_{n-1}-2\theta_{n}\right)+\frac{\tilde{\gamma}_1}{I}\dot{\theta}_n\nonumber\\
&-\frac{\tilde{\gamma}_2}{I}\left(\dot{\theta}_{n+1}+\dot{\theta}_{n-1}-2\dot{\theta}_n\right)+\omega_0^2\sin(\theta_n)\nonumber\\
&+f\omega_d^2\cos(\omega_d t)\cos(\theta_n)=0
\end{align}
where the natural pendulum frequency is given by $\omega_0^2 = (mgl/2 + Mgl)/I$ and the dimensionless forcing coefficient is $f = A\omega_0^2/g$. We use parameter values consistent with the experimental setup of Ref. \cite{CuevasPRL2009}:  $\beta = 0.0165$ Nm/rad, $m = 13$ g, $l = 25.4$ cm, $A = 1.12$ cm, $\tilde{\gamma}_1 = 284$ g cm$^2$/s and $\tilde{\gamma}_2 = 70$ g cm$^2$/s.  The driving frequency $\omega_d$ is an experimental control parameter.  In principle the pendulum bob mass can also be varied, however we take $M = 1.8$ g.  The model can be further simplified by scaling time using the natural frequency, $t \rightarrow t/\omega_0$, to reach the dimensionless form:
\begin{align}
\label{normmod}
\ddot{\theta}_n&- c\left(\theta_{n+1}+\theta_{n-1}-2\theta_{n}\right)+\gamma_1\dot{\theta}_n\nonumber\\
&-\gamma_2\left(\dot{\theta}_{n+1}+\dot{\theta}_{n-1}-2\dot{\theta}_n\right)\nonumber\\
&+\sin(\theta_n)+f\omega^2\cos(\omega t)\cos(\theta_n)=0
\end{align}
where $c = \beta/(I\omega_0^2)$, $\omega = \omega_d/\omega_0$ and $\gamma_i = \tilde{\gamma_i}/(I\omega_0)$.  In this dimensionless model the parameters take the values: $c=0.799$, $f=0.060$, $\gamma_1 = 0.010$, $\gamma_2=0.0024$.  We consider the effect of varying the frequency ratio $\omega$, and to a lesser extent the forcing amplitude $f$ (corresponding to a variation of the lateral driving amplitude $A$ in the physical parameters).

\section{Single pendulum dynamics}
Earlier analysis of a horizontally shaken single pendulum has revealed the possibility of chaotic dynamics \cite{VanDoorenCSF1996} and oscillations about a nonzero equilibrium point \cite{SchmittND1998}.  We are primarily interested in exploring some of the dynamics possible in the experimental set-up of Ref. \cite{CuevasPRL2009}, so we begin by examining the period-1 solutions within the experimentally accessible parameter space.  To this end we consider Eqs. (\ref{normmod}) in the limit of a single pendulum, i.e. $N = 1$, $c=\gamma_2 = 0$.  

\subsection{Period-1 solutions}

In the presence of damping and low amplitude forcing we expect to find regular solutions oscillating with the period of the driving force.  We search for these periodic solutions numerically using the classical shooting method (see, e.g., the discussion in Ref. \cite{VanDoorenCSF1996}), seeking a solution across one forcing period.  The advantage of the shooting method is that the eigenvalues of the correction matrix (used in the shooting iterations) correspond to the Floquet multipliers of the periodic solutions and so determine the stability of the converged solution.  A periodic solution whose Floquet multipliers 
have a magnitude less than or equal to 1 is stable.

\begin{figure}
\centering
 \begin{minipage}{0.95\columnwidth}
  \centering
  \includegraphics[width=0.95\columnwidth]{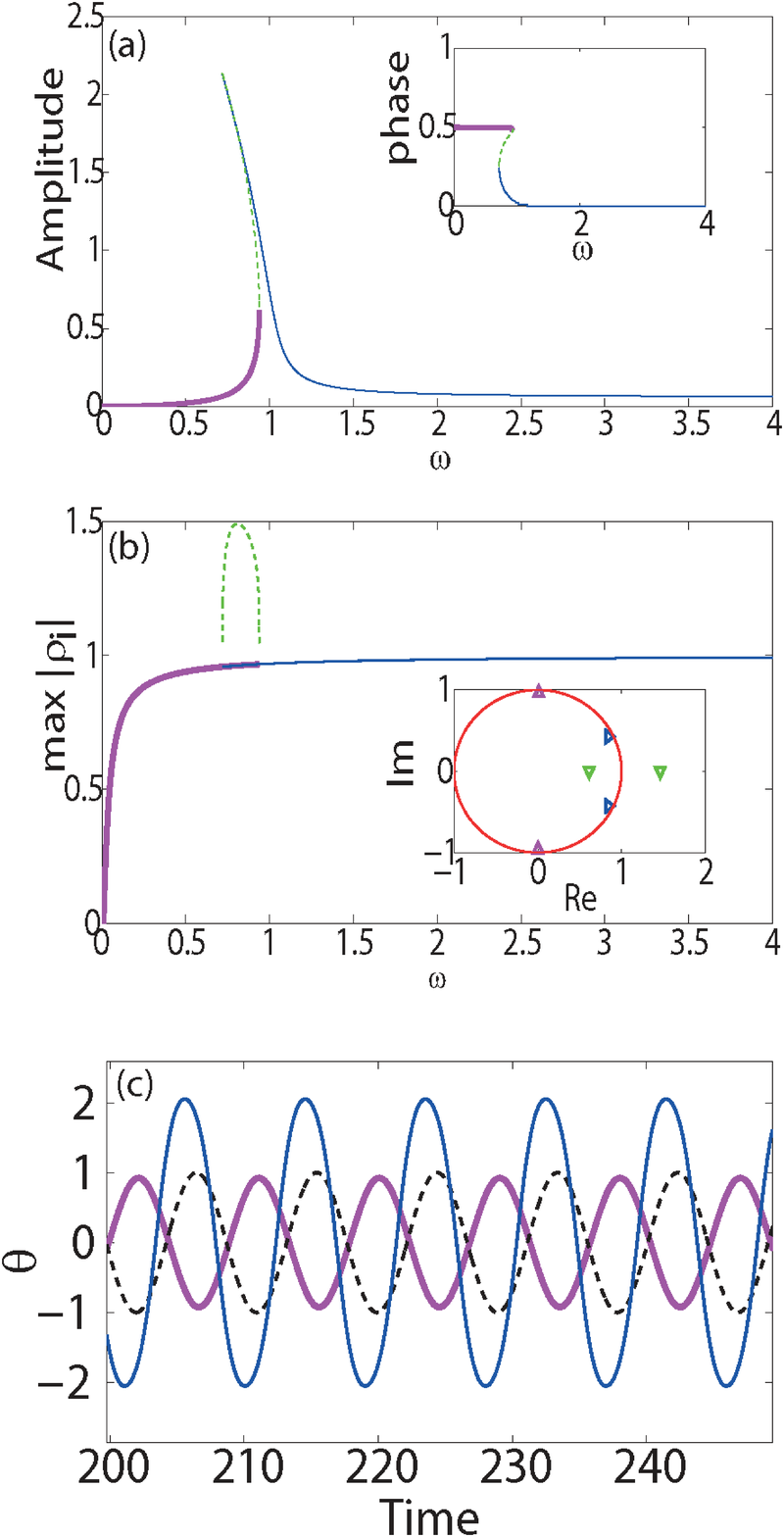}
  \end{minipage}
\caption{(color online).  Families of period-1 solutions in the driven system at low forcing amplitude. $f=0.0597,~\gamma_1=0.01$.  (a) The amplitudes of stable and unstable solutions (solid and dashed lines respectively) as a function of $\omega$ with the inset showing the corresponding phase; (b) the maximum absolute value of the corresponding Floquet multipliers and the inset showing the spectrum corresponding to the three different solutions at $\omega=0.8$ (triangles down: dashed line family; triangles to side: thin solid line family; triangles up: thick solid line family); (c) the appearance of the two stable solutions corresponding
to, respectively, nearly in-phase (thin solid line family in (a) and (b)) and out-of-phase (thick solid line family in (a) and (b)).  For comparison, force is shown as a dashed line.}
\label{singfam1}
\end{figure}

Proceeding with our numerical method we find bistability in a region near the resonance, as is evident in Fig. \ref{singfam1}(a), where we see the familiar foldover effect for a driven sinusoidal pendulum (the resonance drifts to lower frequencies with higher amplitude, due to the softening nature of the sinusoidal nonlinearity).  We see that in the region $\omega = [0.718, 0.9385 ]$, we have two stable solutions, one at high amplitude, the other at low amplitude.  As evident in the inset in Fig. \ref{singfam1}(a) the low amplitude solution is almost exactly out of phase with the force, while the high amplitude solution is closer to being in phase.  The phase here is calculated as: 
\begin{equation}
\label{phase_per1}
{\rm phase} = \frac{\omega}{2\pi}{\rm mod}(t_{\theta_{max}},2\pi/\omega)
\end{equation}
where $t_{\theta_{max}}$ is a time at which the pendulum has maximum amplitude.  This calculation for the phase thus indicates when the pendulum is at maximum 
amplitude relative to the forcing period (for instance a value of 0.5 means it is exactly out of phase with the force).

The stability of the solutions follows from the maximum absolute values of the Floquet multipliers, shown in Fig. \ref{singfam1}(b).  Both the high and low amplitude solutions have magnitudes less than 1 and so are stable.  The inset in Fig. \ref{singfam1}{b} shows details of the spectra at $\omega = 0.8$.  In Fig. \ref{singfam1}(c) we show examples of the stable solutions at $\omega = 0.8$, with the phase and amplitude relationships evident.  

\begin{figure}
\centering
 \begin{minipage}{0.95\columnwidth}
  \centering
  \includegraphics[width=0.95\columnwidth]{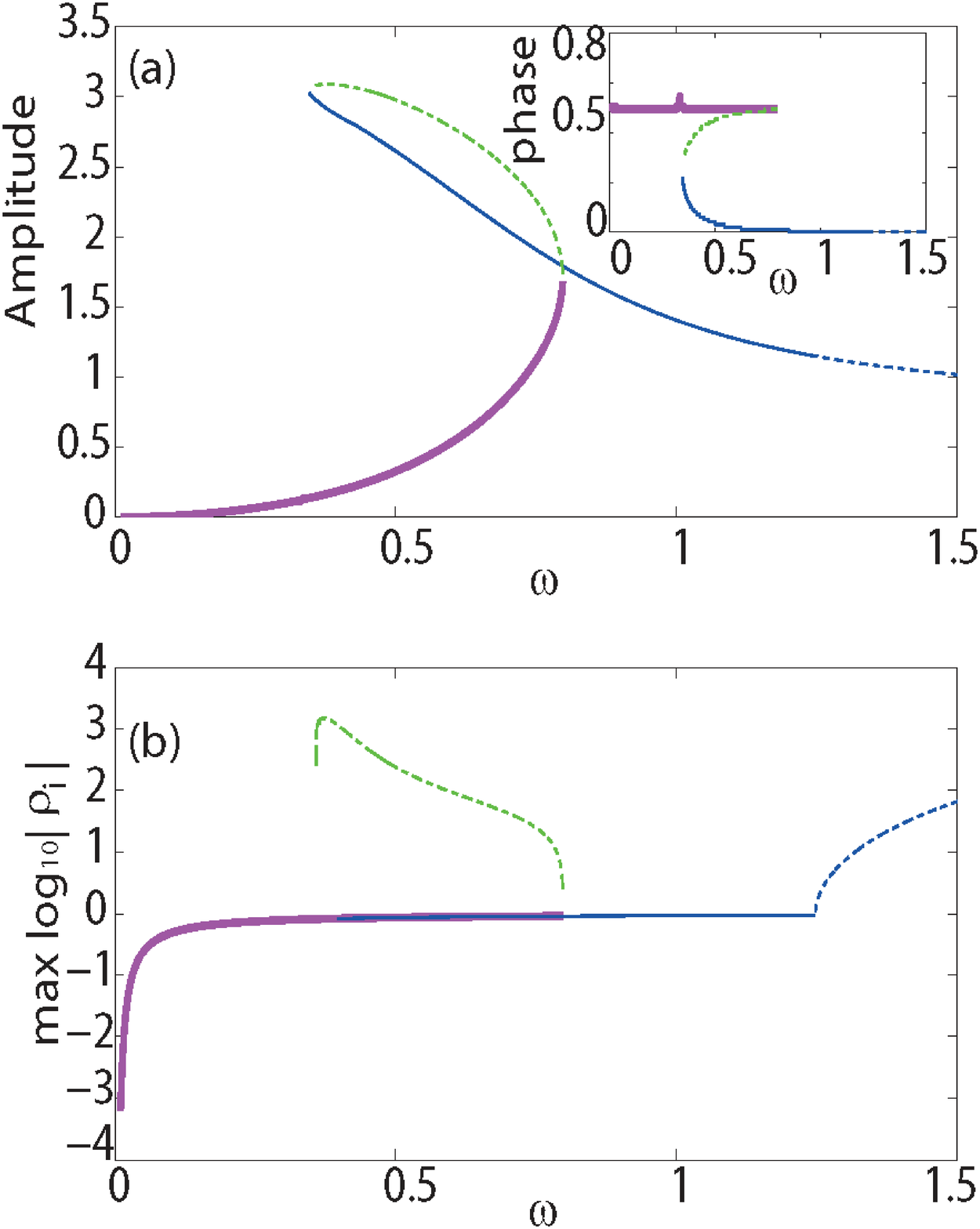}
  \end{minipage}
\caption{(color online).  Family of period-1 solutions in the driven system at large forcing amplitude, $f=1.0,~\gamma_1=0.01$. (a) The amplitudes of stable and unstable solutions (solid and dashed lines respectively) with the inset showing the corresponding phase; (b) the maximum absolute value of the stability 
matrix eigenvalues.}
\label{singfam2}
\end{figure}

While it is not a focus of our work, it is interesting to compare the resonance picture at large forcing amplitude $f = 1$, which has more in common with the analysis of \cite{VanDoorenCSF1996}.  As can be seen in Fig. \ref{singfam2}(a) the amplitude response exhibits a cross-over point, and a significantly more complex spectrum (Fig. \ref{singfam2}(b)).  The phase response (Fig. \ref{singfam2}(a)[inset]) is still much the same as that observed at low forcing amplitude.  
As a  final point,  we note that the high amplitude solution becomes unstable at higher frequency, unlike the low forcing case.

\subsection{Approximate solution}
\label{sec_approx}

Looking ahead to our multi-pendulum analysis we seek an approximate solution for the period-1 response.  We find that we can obtain a good prediction of this response by simply assuming the solution takes the harmonic form: 
\begin{equation}
\label{ansatz}
\theta = V_{c} \cos(\omega t+\phi), 
\end{equation}
where $V_c$ is the amplitude of the pendulum oscillation and $\phi$ is the phase offset from the driving phase $\omega t$.  To proceed, we make two assumptions which allow us to simplify the problem.  Firstly, we simplify our equation of motion Eq. (\ref{normmod}) by replacing the trigonometric nonlinearities with their low order algebraic expansions, i.e. $\sin\theta \approx \theta-\frac{1}{6}\theta^{3}$ and $\cos\theta \approx 1-\frac{1}{2}\theta^{2}$.  This leads to the new (approximate) equation of motion:

\begin{equation}
	\ddot{\theta}+\gamma_1\dot{\theta}+(\theta-\frac{1}{6}\theta^{3})+f\omega^2\cos(\omega t)(1-\frac{1}{2}\theta^{2})=0
\label{simplifiedmodel}
\end{equation}

Secondly, remembering the trigonometric identity $\cos^3(\omega t+\phi) = \frac{3}{4}\cos(\omega t + \phi) + \frac{1}{4}\cos(3(\omega t + \phi))$, we assume all contributions not at the frequency of the driving force are weak and can be neglected (the so-called rotating wave approximation).  Incorporating this assumption into our simplified model (\ref{simplifiedmodel}) we obtain two equations for $(V_c,\phi)$ by multiplying by $\cos(\omega t)$ and $\sin(\omega t)$ respectively and integrating out the time dependence (integration over $[0,2\pi/\omega]$):

\begin{align}
&-\frac{\pi V_c}{8\omega}\Bigg( 2\omega^2fV_c\cos(\phi)^2+f\omega^2V_c-8f\omega^2+8\gamma_1\omega \sin(\phi)\notag\\  &+8\omega^2\cos(\phi)-8\omega_0^2\cos(\phi)+V_c^2\omega_0^2\cos(\phi)\Bigg) =0
\label{eq1Vphi} \\
&\frac{\pi V_c}{8\omega}\Bigg( 8\omega^2\sin(\phi)-8\omega_0^2\sin(\phi)+2\omega^2 f V_c\cos(\phi)\notag\\  &+V_c^2\omega_0^2\sin(\phi)-8\gamma_1\omega\cos(\phi)\Bigg) =0
\label{eq2Vphi}
\end{align} 

\begin{figure}[htbp]
\includegraphics[width=0.9\columnwidth]{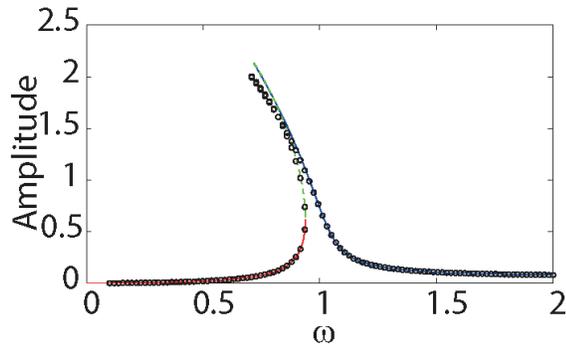}
\caption{(color online).  Comparison between single-pendulum period-1 amplitudes found numerically (lines), and using the ansatz (\ref{ansatz}), (circles), for $f=0.0597,~\gamma_1=0.001$.}
\label{comparison}
\end{figure}

We solve these equations using the symbolic mathematics package ${\rm Maple}^{\texttrademark}$, after setting values for $\gamma_1$ and $\omega$.  The results are shown in Fig. \ref{comparison} with $\gamma_1 = 0.01$ and $\omega$ varying, showing good qualitative agreement with the numerical results, even when $V_c$ is large.

\subsection{Dynamical response}

\begin{figure}[htbp]
\includegraphics[width=0.95\columnwidth]{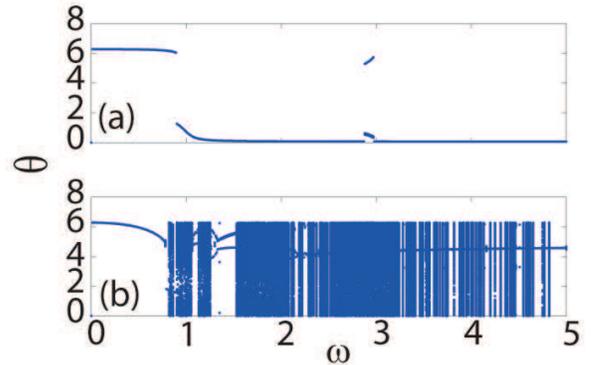}
\caption{Bifurcation plot of $\theta$ vs $\omega$ constructed from a fixed initial condition $\theta = 0.5$, $\dot{\theta} = 0$, where $\theta(t)$ is plotted every $t = 2\pi/\omega$ from $t = 4000$ to $t = 5000$. (a) $f = 0.059$ showing appearance of period-1 and multi-period solutions and (b) $f = 1.0$ showing predominantly aperiodic dynamics.}
\label{bif}
\end{figure}

We turn now to the dynamical response of the single pendulum with the non-zero initial condition $\theta(0) = 0.5$ and $\dot{\theta}(0) = 0$.  We represent the dynamical response as a bifurcation plot, plotting all values of $\theta$ sampled at the forcing frequency (between $t = 4000$ and $t = 5000$ to allow initial transients to die out), for a given forcing frequency $\omega$.  A single value of $\theta$ indicates a period-1 solution.  Multiple values for a given $\omega$ indicate a longer period solution.  At the low forcing amplitude of $f = 0.0597$ we see in Fig. \ref{bif}(a) a relatively simple plot, with the generation of a single period solution from the given initial condition for most of the frequency values, except around $\omega= 3$ where it appears the dynamics has converged to a multi-period solution.  Note that this has only appeared in the dynamics because we have chosen a non-zero initial condition.  For small (or zero) initial amplitudes we find instead the dynamics converge to the low amplitude solution.  The jump around $\omega = 0.9$ is due to the change in convergence from the low amplitude solution to the high amplitude solution (thus the jump indicates the presence of bistability). 

By way of contrast we consider also the dynamics at large forcing amplitude $f = 1.0$ in Fig. \ref{bif}(b), where we see the appearance of aperiodic solutions, which we surmise to be chaotic, for large ranges of $\omega$.  A deeper analysis of the nature of the dynamics in this regime, and the possible routes to chaos, is beyond the scope of this work.

\subsection{Period-3 solutions}
\begin{figure}[htbp]
\includegraphics[width=0.95\columnwidth]{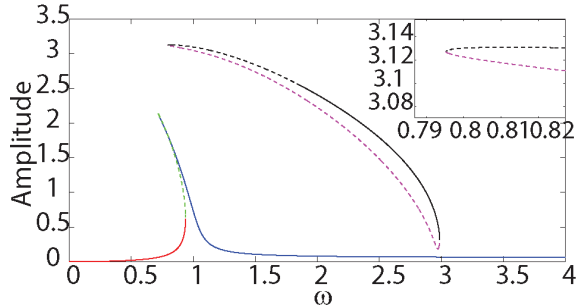}
\caption{(color online).  Period-3 solutions (isola) compared with period-1 solutions (fold-over). The dashed and solid lines correspond to unstable and stable solutions respectively. [Inset] Zoom in showing the left-hand bifurcation point.}
\label{tripsol}
\end{figure}

We find that the jump observed near $\omega = 3$ in Fig. \ref{bif}(a) is due to the existence of a stable three-period solution.  This subharmonic response is well known in the pendulum system (see e.g. \cite{ChesterJIMA1975}, or for a textbook treatment \cite{Nayfeh1979}).  To examine in more detail this solution we turn again to our numerical shooting method but this time seek solutions with period $T=\frac{6\pi}{\omega}$.  We find the existence of a saddle-node bifurcation near $\omega = 3$ and the appearance of lower amplitude (unstable) and higher amplitude (stable) solutions, as can be seen in Fig. \ref{tripsol}.  
Perhaps somewhat surprisingly the stability region in driving frequency for the upper branch solution is large, extending to less than $\omega = 2$.    The two solutions are of much larger amplitude than the period-1 solutions, and appear to form an isola (as seen in \cite{ChesterJIMA1975}), well separated from the low amplitude solution.  
 
\section{Multi-pendulum dynamics}

We now turn to the multi-pendulum case, i.e. the system (\ref{normmod}) with $N>1$.  We begin by examining the fundamental oscillation mode (all pendulums synchronised) and find similar results to those of the single-pendulum case, albeit with the emergence of instability for $N \ge N_{cr}$.  We then turn our attention to the nature of the breather solutions (localized energy states), examine their connection to the fundamental mode solution and explore methods for their generation.
  
\subsection{Fundamental oscillation mode}
\label{sec:numerical}

\begin{figure}[htbp]
\centering
 \begin{minipage}{0.95\columnwidth}
  \centering
  \includegraphics[width=0.95\columnwidth]{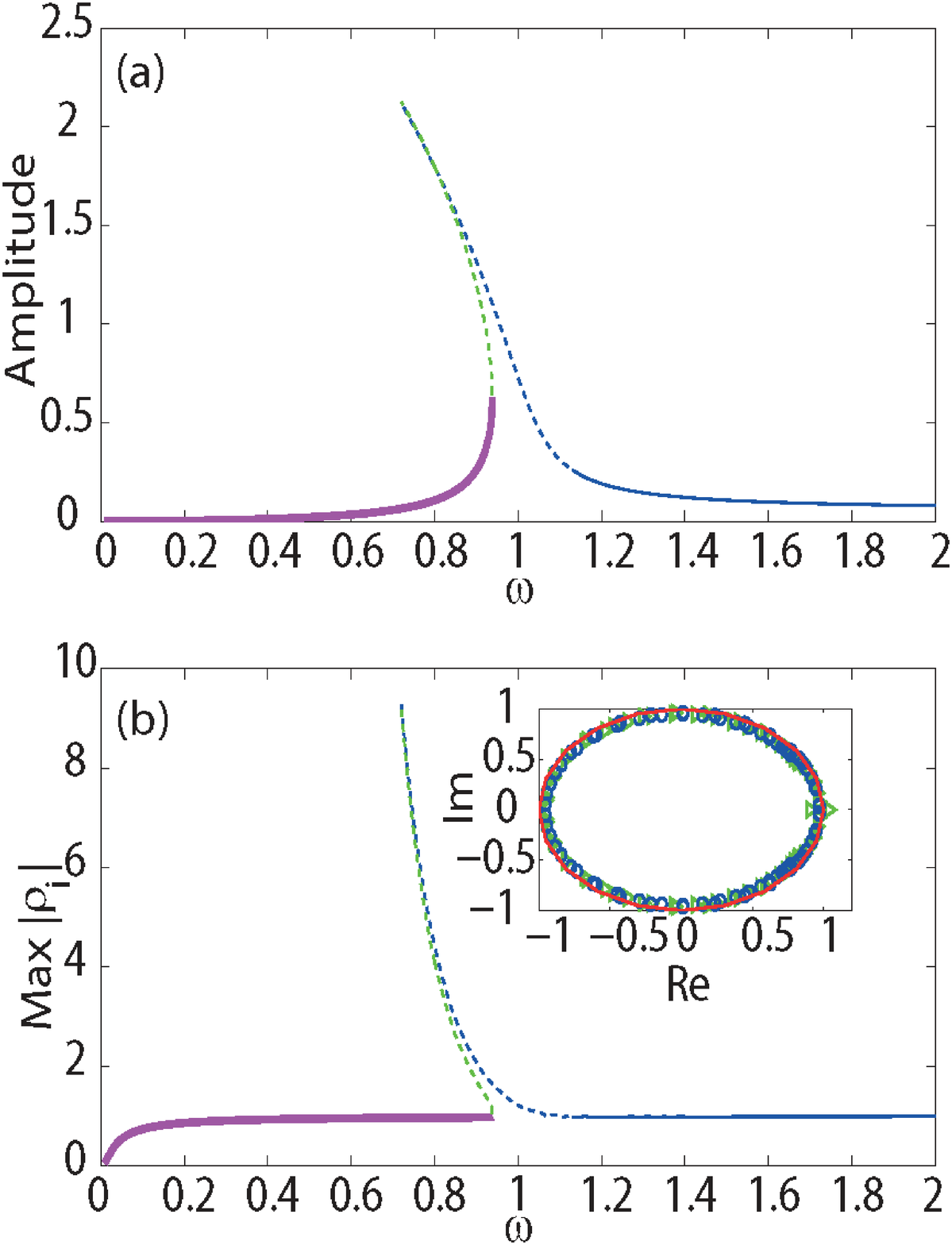}
  \end{minipage}
\caption{(color online).  Family of period-1 solutions in the driven system at low forcing amplitude for $N = 41$. $f=0.0597,~\gamma_1=0.001, ~\gamma_2=0.020,~c=0.799$ showing (a) amplitude response and (b) associated maximum instability eigenvalue with [inset] spectrum on either side of the stability change ($\omega = 1.06$ and $\omega = 1.07$) for the high amplitude solution.}
\label{multifam1}
\end{figure}

We find that the period-1 solutions mirror those found in the single-pendulum case, except for the appearance of instability in the large amplitude solution beyond a critical chain length.  We show in Fig. \ref{multifam1} the dependence of the period-1 solutions on $\omega$ for $N = 41$. The solid lines indicate stable solutions and the dashed lines indicate unstable ones.  The instability in this case appears at $\omega = 1.07$.  As can be seen in the inset of Fig. \ref{multifam1}(b) the instability appears through the collision of complex eigenvalues producing real eigenvalues with magnitude greater than 1.

\begin{figure}[htbp]
\includegraphics[width=0.95\columnwidth,height=0.20\textheight]{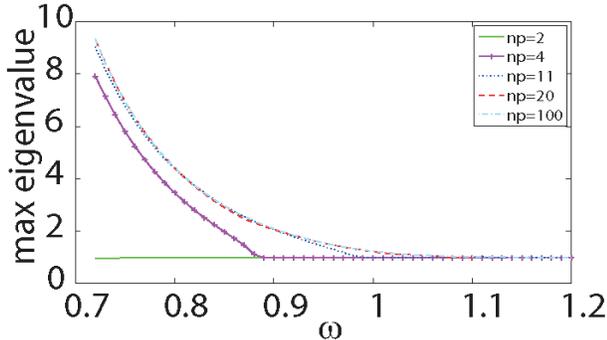}
\caption{(color online).  Appearance of maximum instability eigenvalue for the period-1 solution at different chain lengths, showing the appearance of a critical chain length for the appearance of instability.  Also evident is a critical value of driving frequency $\omega$ beyond which the period-1 solution appears to be stable for all chain lengths.}
\label{instabilityregion}
\end{figure}

In Fig. \ref{instabilityregion} we compare the results for chains with different numbers of pendulums, and find the critical chain length for the appearance of the instability is $N_{cr} = 4$.  Also evident in Fig. \ref{instabilityregion} is a suggestion of convergence to a maximum $\omega$ beyond which we have stability, even for longer chains.  The instability threshold for $N = 20$ is $\omega = 1.07$, while for $N = 100$ it is $\omega=1.14$.  We should note that this analysis is focused only on the solutions for which all pendulums behave identically.  The full bifurcation picture is expected to be significantly more complicated (as for instance seen in a recent work considering coupling between two Duffing oscillators \cite{IkedaJCND2013}).  The detailed bifurcation study is beyond the scope of this work, instead we study the nature of the instability development for the fundamental oscillation mode at long chain lengths, through modulational instability analysis.

\subsection{Modulational Instability Analysis}

We look for the emergence of modulational instability (MI) in the oscillator chain, using standard MI analysis, as for instance carried out for chains of periodically-forced anharmonic oscillators \cite{BurlakovPRL1998}.

We begin with the approximate solution $\theta_n = V_c \cos(\omega t + \phi)$, equivalent to the single pendulum solution calculated in Section \ref{sec_approx}.  We then add a perturbation to this solution in the form:
\begin{align}
\delta\theta(n) &= \frac{1}{2}\cos(qn)\Bigg( V_{p1}\exp[i(\omega-\Omega)t] \\\nonumber  
&+V_{p2}\exp[i(-\omega-\Omega)t]\Bigg)
\nonumber
\\
&+ C.C.
\label{perturbation}
\end{align}
where $V_{pj}$ is the complex amplitude, and $q$ and $\Omega$ are 
the vector and (in principle) complex frequency of the perturbation, respectively.  Substituting $\theta_n + \delta\theta_n$ into (\ref{normmod}) and keeping only terms to first order in the perturbation (\ref{perturbation}), we obtain an equation for $V_{p1}$ and for $V_{p2}$.  Multiplying these two equations together and integrating over a single forcing period $[0,2\pi/\omega]$ to remove the time dependence we end up with the following equation relating the instability wavenumber $q$ and growth rate $\Omega$:
\begin{align}
\Bigg[ &-(\omega-\Omega)^2+ i\gamma_1(\omega-\Omega)-c(2\cos(q)-2)\\\notag
&+\omega_0^2(\omega-\Omega)-2i\gamma_2(\cos(q)-1)-\frac{1}{4}\omega_0^2 V_c^2 \Bigg]\\\notag
&\times \Bigg[ -(-\omega-\Omega)^2-i\gamma_1(\omega+\Omega)-c(2\cos(q)-2)\\\notag &+\omega_0^2+2i\gamma_2(\cos(q)-1)(\omega+\Omega)-\frac{1}{4}\omega_0^2 V_c^2 \Bigg]\\\notag &=\frac{1}{32} \Bigg( \omega_0^4 V_c^2+4\omega_0^2V_c f\omega^2\cos(\phi) \\\notag &+4f^2\omega^4+8f^2\omega^2\cos(\phi)^2 \Bigg)
\label{MIeq}
\end{align}
To proceed, we substitute the $V_c$ and $\phi$ associated with the fundamental mode of interest at a given $\omega$ (found from solving Eqs.(\ref{eq1Vphi}) and (\ref{eq2Vphi})), and solve for $\Omega$ at a given $q$.  We see from the nature of the perturbation (\ref{perturbation}) that for any $\Omega$ with a positive imaginary part we will have growth of the perturbation, and a resultant instability with wavenumber given by $q$.

\begin{figure}[htbp]
\includegraphics[width=0.9\columnwidth]{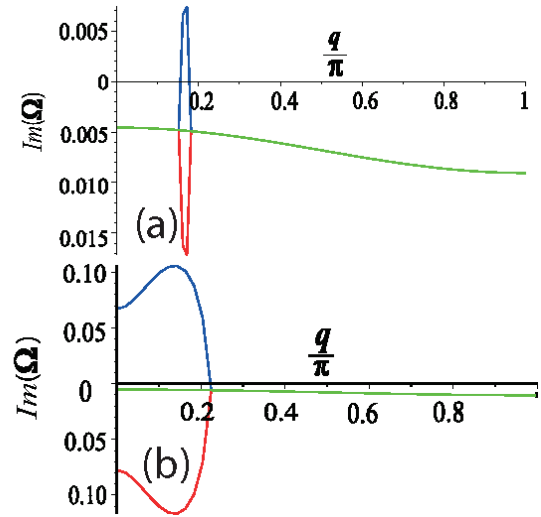}
\caption{(color online).  (a) The maximal growth
rates associated with the modulational stability analysis
when (a) $\omega=1.09$  and (b) $\omega=0.95$.}
\label{qOm_unstable}
\end{figure}

\begin{figure}[htbp]
\includegraphics[width=0.9\columnwidth]{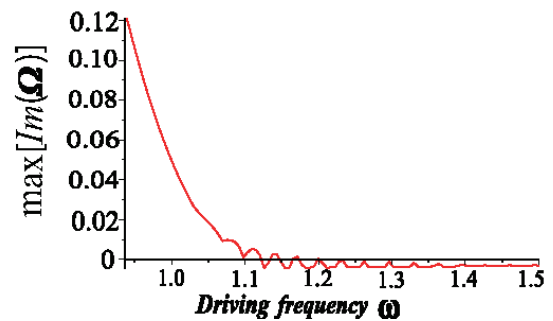}
\caption{(color online).  The maximum imaginary part of $\Omega$ variation
as a function of the driving frequency $\omega$, showing extended stability beyond $\omega > 1.15$.}
\label{mapleinstab}
\end{figure}

As we can see in Fig. \ref{qOm_unstable} for $\omega = 1.09$ (top panel) 
we find instability for a small range of wavenumbers around $q = 0.165\pi$.  This suggests the instability development will progress with a characteristic periodicity corresponding to about 12 pendula.  As we move deeper into the unstable region we see in Fig. \ref{qOm_unstable} for $\omega= 0.95$ (bottom
panel) that the interval of unstable wavenumbers increases, as well as the instability growth rate.  Examining the dependence of the maximum growth rate $\operatorname{Im}\Omega$ versus $\omega$ we see in Fig. \ref{mapleinstab} some oscillations around the critical threshold, but extended stability above $\omega = 1.15$, agreeing well with the numerical prediction of $\omega = 1.14$ shown in Fig. \ref{instabilityregion} for 100 pendulums.

\subsection{Instability dynamics}

We turn now to an investigation of the time-dependent response of the pendulum chain, i.e. evolution of system (\ref{normmod}) given initial conditions for each pendulum amplitude and velocity.  While there are some interesting questions concerning the basins of attraction of the various periodic solutions we have found, we shall consider here only the dynamics following from the initial conditions $\theta_n(0) = \dot{\theta}_n(0) = 0$.

\begin{figure}[htbp]
\centering
 \begin{minipage}{1.0\columnwidth}
  \centering
  \includegraphics[width=1.0\columnwidth]{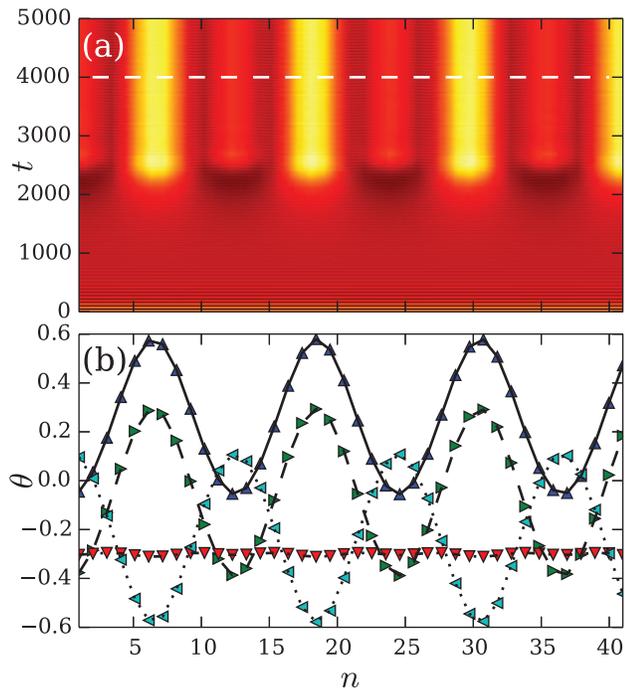}
  \end{minipage}
  \caption{(color online). Development of modulational instability at $\omega = 1.09$ for a chain of 41 pendulums. (a) Plot of total energy per pendulum (black: -1, white: -0.8), showing formation of regular high energy peaks. (b) Amplitude at dashed line in (a), given by $t = 4001$ (up triangles), $t = 4002$ (right triangles), $t = 4003$ (down triangles) and $t=4004$ (left triangles) showing larger oscillations at high energy areas. Lines through the pendulum amplitudes are shown to aid the eye.  Initial conditions $\theta_n = 0$, $\dot{\theta}_n = 0$}
  \label{mi_dynom1p09}
  \end{figure}

Firstly, we see in Fig. \ref{mi_dynom1p09} for $\omega = 1.09$ the spontaneous symmetry breaking accompanying the appearance of modulational instability.  We monitor the energy per pendulum to visualize the dynamics, with the 
associated energy 
density given by:
\begin{equation}
\label{energy}
E_n = \frac{1}{2}\dot{\theta}_n^2 - \cos(\theta_n) + c\frac{1}{2}\left[(\theta_{n+1} - \theta_n)^2 + (\theta_{n} - \theta_{n-1})^2\right]
\end{equation}
Fig. \ref{mi_dynom1p09}(a) shows the energy per pendulum as a function of time $t$ and pendulum number $n$ with white corresponding to $E_n = -0.8$ and black $E_n = -1$.  As can be seen after some initial transient time the symmetry is spontaneously broken through energy localization.  We can see in Fig. \ref{mi_dynom1p09}(b) that the periodicity of the emergent pattern is roughly 12 pendula, as 
was earlier predicted by our modulational instability analysis.

\begin{figure}[htbp]
\centering
 \begin{minipage}{1.0\columnwidth}
  \centering
  \includegraphics[width=1.0\columnwidth]{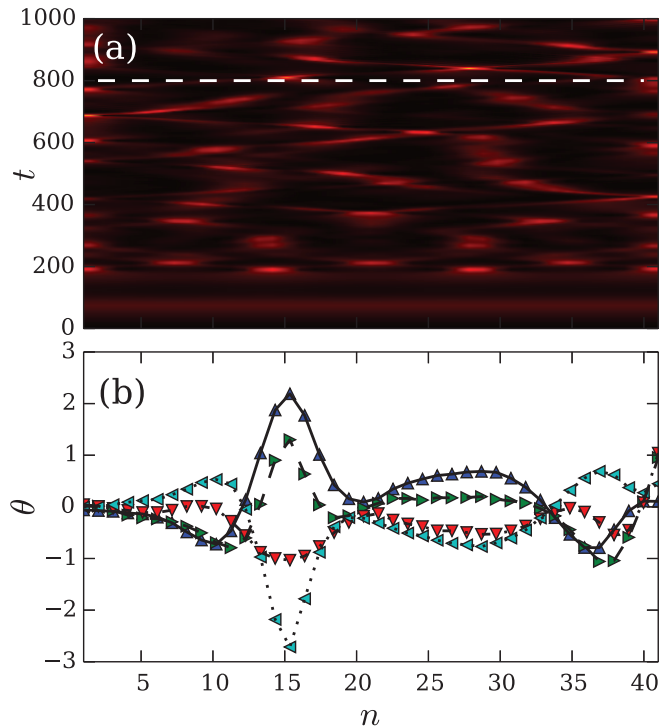}
  \end{minipage}
  \caption{(color online). Rapid development of instability at $\omega = 0.95$ for a chain of 41 pendulums. (a) Plot of total energy per pendulum (black: -1, white: 7), showing formation of high energy excitations. (b) Amplitude at dashed line in (a), given by $t = 800$ (up triangles), $t = 801$ (right triangles), $t = 802$ (down triangles) and $t=803$ (left triangles) showing a transient localized excitation. Lines through the pendulum amplitudes are shown to aid the eye.  Initial conditions $\theta_n = 0$, $\dot{\theta}_n = 0$}
  \label{mi_dynom0p95}
  \end{figure}

Our MI analysis also showed that decreasing the driving frequency increases both the instability growth rate and the range of unstable wavenumbers.  To explore the effect of this in the dynamics we consider $\omega = 0.95$ with 101 pendulums.  As can be seen in Fig. \ref{mi_dynom0p95} the instability appears much more rapidly (in comparison with Fig.~\ref{mi_dynom1p09}), and without a clearly dominant wavenumber to the instability, as may be expected from the wide interval
of unstable wavenumbers, including many with roughly similar growth rates.  
Instead, we see highly nonstationary dynamics and apparent energy localization.  This, in turn, motivates us to study in Section \ref{breather} the localized states spontaneously arising in the instability evolution.

\begin{figure}[htbp]
\centering
 \begin{minipage}{1.0\columnwidth}
  \centering
  \includegraphics[width=1.0\columnwidth]{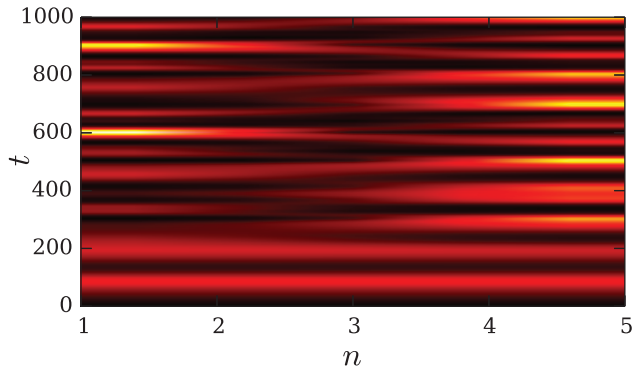}
  \end{minipage}
  \caption{(color online). Instability development in a short chain of five pendulums at $\omega = 0.94$.  No stationary pattern is evident, instead energy appears to move through the chain and become temporarily trapped at the end pendulums.  The initial conditions are $\theta_n = 0$, $\dot{\theta}_n = 0$.  Colormap indicates total energy per pendulum (white: 1.76, black: -1)}
  \label{MInpen5}
  \end{figure}

Finally we consider the emergence of instability in short chain lengths, below the 12 pendulum periodicity predicted by our MI analysis.  Our earlier analysis of the pendulum chain indicated that the large amplitude state becomes unstable when consisting of four pendulums, for frequencies below approximately $\omega = 0.9$ (see Fig. \ref{instabilityregion}).  In practice at this driving frequency we find the pendulum chain will converge to the stable low amplitude solution (which exists up until approximately $\omega = 0.938$).  The instability region increases with chain length and in fact already a chain of five pendulums is unstable beyond the low amplitude existence region.  We see in Fig. \ref{MInpen5} the instability dynamics which emerge at $\omega = 0.94$.  The instability pattern of the long chain is gone and instead we see that the energy appears to move through the chain becoming temporarily trapped at the ends. It is important to
note that the modulational stability analysis above applies to the infinite chain, since the wavenumber parameter $q$ was taken to be a continuous variable.  An interesting direction for future work would be to consider separately the case of short chains, with $q$ suitably quantized.

\subsection{Period-1 Breather solutions}
\label{breather}

\begin{figure}[htbp]
\centering
 \begin{minipage}{0.95\columnwidth}
  \centering
  \includegraphics[width=0.95\columnwidth]{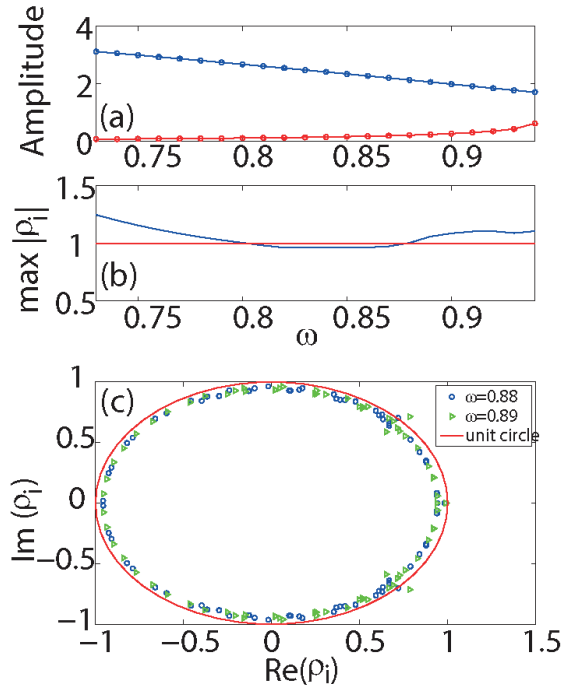}
  \end{minipage}
\caption{(color online).  Family of onsite breathers for $N = 41$. (a) amplitude of central site (blue) and tails (red); (b) maximum eigenvalue; (c) the eigenvalues from stable and unstable breathers in the complex plane.}
\label{breather_onsite}
\end{figure}

\begin{figure}[htbp]
\centering
 \begin{minipage}{0.95\columnwidth}
  \centering
  \includegraphics[width=0.95\columnwidth]{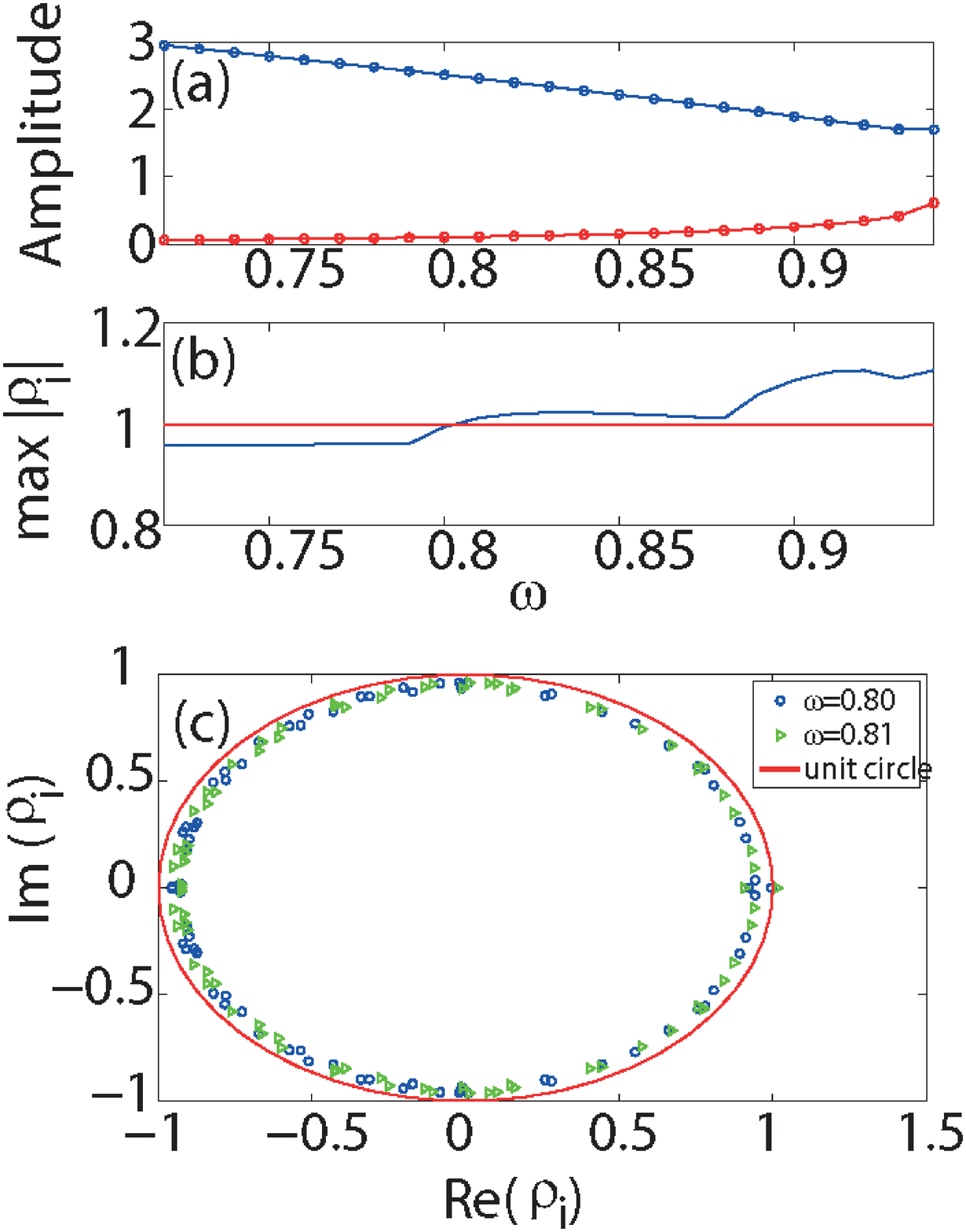}
  \end{minipage}
\caption{(color online). Family of off-site breathers for $N = 41$. (a) amplitude of central site (blue) and tails (red); (b) maximum eigenvalue; (c) the eigenvalues from stable and unstable breathers in the complex plane.}
\label{breather_offsite}
\end{figure}

The energy localization we have observed in our instability dynamics appears 
somewhat connected to the breather solutions observed in Ref. \cite{CuevasPRL2009}.  Thus, we continue our analysis by examining this problem, namely the
form and stability of the period-1 breather solutions.  In agreement with the results of Ref. \cite{CuevasPRL2009} we find two families of breather solutions corresponding to on-site (Fig. \ref{breather_onsite}) and off-site (Fig. \ref{breather_offsite}) configurations.  We find these solutions by beginning in the anti-continuous limit \cite{MarinN1996}, with the central site(s) corresponding to the high amplitude single pendulum solution and outer sites the low amplitude solution.  Both families are only stable for a limited, and (roughly) 
mutually exclusive, range of driving frequencies $\omega$, with the onsite 
state being stable at higher frequencies.  

Examining the Floquet multipliers at the instability crossing we see that the 
on-site breather becomes unstable due to the growth of a complex conjugate pair (Fig. \ref{breather_onsite}(c)), while the off-site breather becomes unstable through the growth of a purely real Floquet multiplier (Fig. \ref{breather_offsite}(c)).

\begin{figure}[htbp]
\includegraphics[width=0.9\columnwidth]{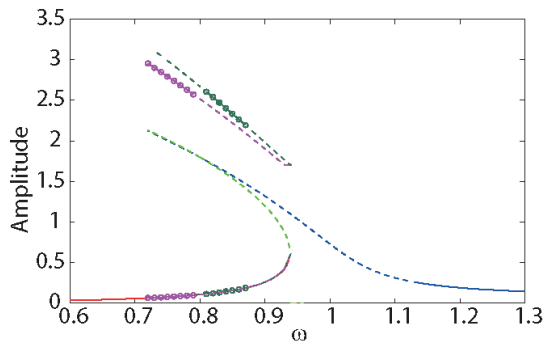}
\caption{(color online).  Breather solution for $N = 41$ compared with period-1 solutions. The green and magenta lines stand for the on-site and off-site breather respectively, and the circles mean stable states.}
\label{breathervsfamily}
\end{figure}

The small range of $\omega$ for which the solutions exist is directly related to the region of bistability in the single pendulum case.  This can be most clearly seen by superimposing a plot of the amplitudes of the breather maximum and tails on the single pendulum amplitude response shown originally in Fig. \ref{singfam1}(a).  We show this in Fig. \ref{breathervsfamily} where the maximum amplitudes of the central site(s) for the two breather families (upper dashed lines, top: on-site, lower:off-site) is shown along with the regions over which these families are stable (upper circles) and the amplitudes of the breather tails (lower circles), superimposed on the pendulum chain results (lower solid and dashed lines).  It is evident in this figure that the breathers only exist in the region of bistability for the period-1 solutions of the pendulum chain.  Interestingly there is a region in $\omega$ approximately $[0.94, 1.12]$ where we find no stable solutions.  We will exploit this fact in our breather generation scheme discussed in Section \ref{generation}.

\begin{figure}[htbp]
\centering
 \begin{minipage}{0.95\columnwidth}
  \centering
  \includegraphics[width=0.95\columnwidth]{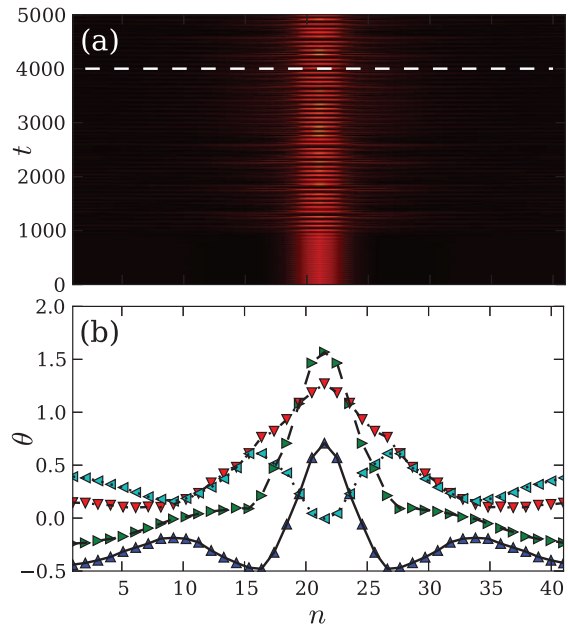}
  \end{minipage}
  \caption{(color online). Dynamics of unstable on-site breather at $\omega = 0.9$ for a chain of 41 pendula. (a) Plot of total energy per pendulum (black: -1, white: 4.8), showing large amplitude oscillations of central sites. (b) Amplitude at dashed line in (a), given by $t = 4001$ (up triangles), $t = 4002$ (right triangles), $t = 4003$ (down triangles) and $t=4004$ (left triangles) showing presence of large amplitude oscillations in breather tails.}
  \label{breathinstab}
  \end{figure}

We finish this section by briefly examining the instability dynamics of an unstable breather.  As an example we consider the on-site breather at $\omega = 0.9$.  For this driving frequency the instability growth rate is weak, and we expect still connected with complex conjugate Floquet multipliers.  The resultant break-up dynamics of the breather shown in Fig. \ref{breathinstab} is indeed slow, showing amplitude oscillations and gradual emission of energy into the surrounding pendulum chain.  By $t= 4000$ the tails of the breather are showing significant amplitude oscillations (see Fig. \ref{breathinstab}(b)), yet the energy is still predominantly localized about the central site of the pendulum chain, indicating the weak nature of the instability.

\subsection{Mixed-frequency breather solutions}

\begin{figure}[htbp]
\centering
 \begin{minipage}{0.95\columnwidth}
  \centering
  \includegraphics[width=0.95\columnwidth]{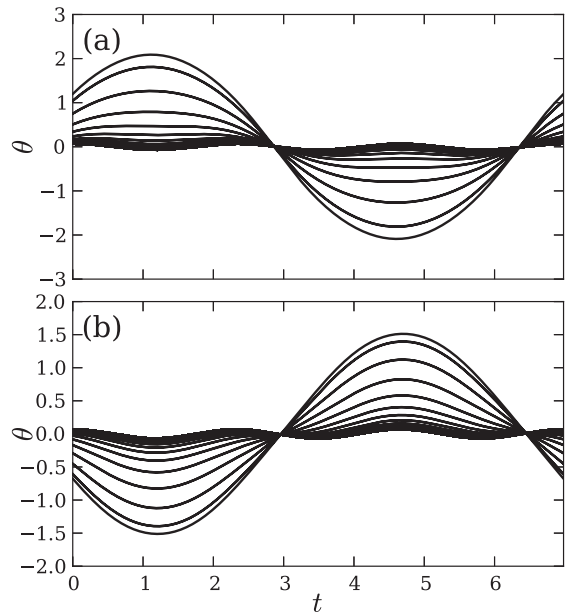}
  \end{minipage}
  \caption{Mixed-frequency breather trajectories across one period ($T = 6\pi/\omega$) for every pendulum ($N = 101$) with $\omega=2.7$. (a) Central pendulums moving at three times the period of the driving force, and out of phase with pendulums at edges (weakly unstable); (b) as above, except central pendulums in phase with tails (strongly unstable).}
  \label{mperbreath}
  \end{figure}

Inspired by the connection between the period-1 single-pendulum solutions and the nature of the period-1 breathers we also examine the possibility of breathers connected to the period-3 single-pendulum solutions.  In particular, the possibility of breathers in the bistable region of low amplitude period-1 solutions and high amplitude period-3 solutions.  We find that indeed breathers exist in this bistable region, with the tails connected to the period-1 solution and the central site(s) connected to the period-3 solution (see Fig. \ref{mperbreath}).  

\begin{figure}[htbp]
\centering
 \begin{minipage}{1.0\columnwidth}
  \centering
  \includegraphics[width=1.0\columnwidth]{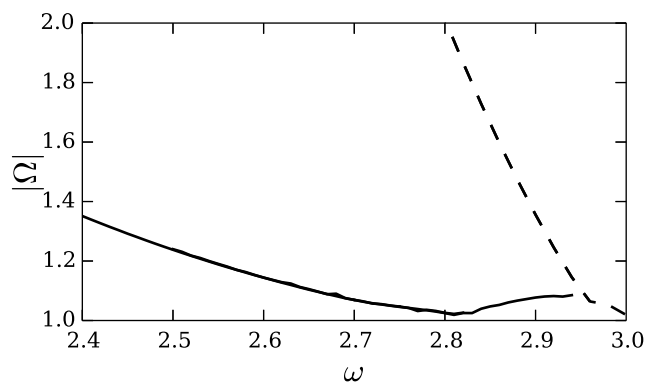}
  \end{minipage}
  \caption{Absolute value of maximum Floquet multiplier for multi-frequency breather solutions in a chain of 101 pendula.  Solid line corresponds to the out-of-phase solution, showing weak instability, while the dashed line corresponds to the in-phase solution showing strong instability.}
  \label{mperfam}
  \end{figure}

Due to their mixed frequency nature more breather families become possible.  In particular, we find that the central sites (moving with one third the frequency of the driving force) may move either out of phase or in phase with the low amplitude tails (moving with the same frequency as the driving force).  These two classes are shown in Fig. \ref{mperbreath}(a) and \ref{mperbreath}(b) respectively, where we have plotted $\theta(t)$ for each pendulum over three forcing periods, for $\omega = 2.7$ and $N = 101$.  Such, so-called subharmonic breathers
have also been found in the context of electrical lattices 
recently in~\cite{lars2}.
Examining the Floquet spectrum for these families we find that the in-phase state is highly unstable (dashed line in Fig. \ref{mperfam}) while the out-of-phase state is only weakly unstable (solid line in Fig. \ref{mperfam}).  

\begin{figure}[htbp]
\centering
 \begin{minipage}{0.95\columnwidth}
  \centering
  \includegraphics[width=0.95\columnwidth]{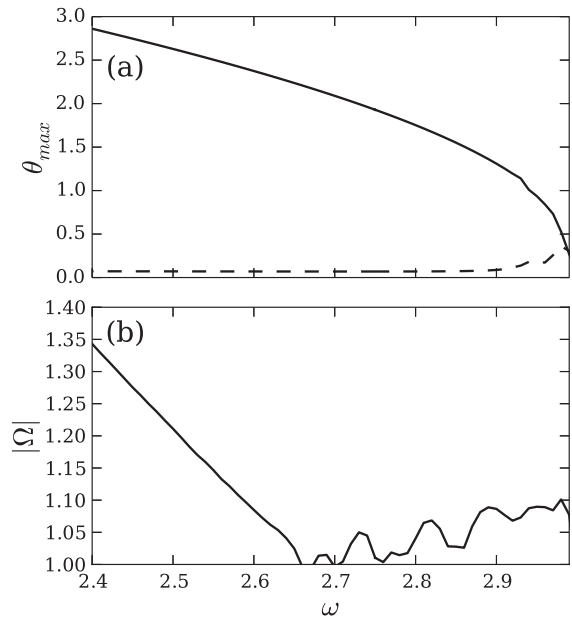}
  \end{minipage}
  \caption{Multi-frequency breather solutions for $N=41$. (a) Central site amplitude (solid line) and edge site amplitude (dashed line) versus $\omega$ for on-site, out-of-phase, breather solution; (b) Maximum absolute value of the instability eigenvalues associated with solutions in (a) showing small windows of stability near $\omega = 2.7$.}
  \label{mperfamN41}
  \end{figure}

While we find no stable examples of multi-frequency breathers for a chain of 101 pendulums, at shorter chain lengths windows of stability appear.  We show in Fig. \ref{mperfamN41} the dependence on driving frequency $\omega$ for the family of on-site out-of-phase mixed-frequency breathers in a chain of 41 pendulums.  The central site amplitude (solid line) and edge site amplitude (dashed line) are shown in Fig. \ref{mperfamN41}(a), with the largest Floquet multiplier shown in Fig. \ref{mperfamN41}(b).  Around $\omega = 2.7$ we see that this mixed-frequency breather is predicted to be stable. As we approach $\omega = 3$ the mixed-frequency breather becomes more and more delocalized, evident in the apparent collision of the central site and tail amplitudes.  We also see some evidence of symmetry breaking in the tail oscillators, which is perhaps the origin of the small loss in continuity of the breather family amplitude near cut-off.  The full analysis of the properties of this breather state are beyond the scope of this work, but would be an interesting direction for future work.
 
\begin{figure}[htbp]
\centering
 \begin{minipage}{0.94\columnwidth}
  \centering
  \includegraphics[width=0.94\columnwidth]{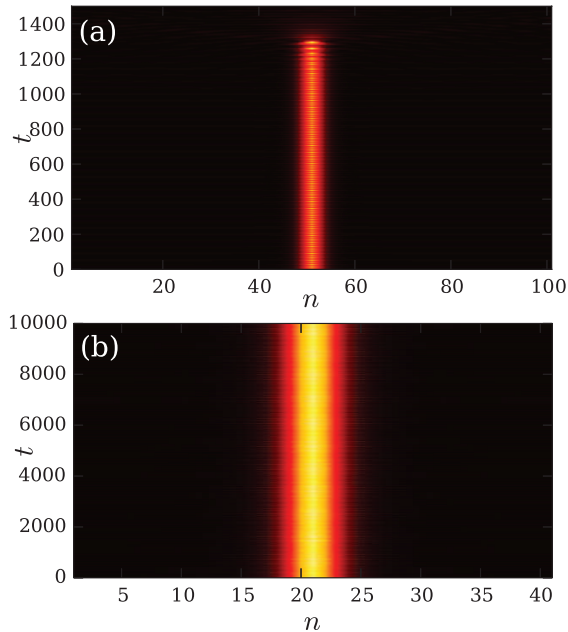}
  \end{minipage}
  \caption{(color online).  Dynamics of multi-frequency breathers for different chain lengths. (a) Weakly unstable case for $\omega = 2.7$ and $N = 101$ showing disappearance of breather (black: -1, white: 2.5); (b) Very long lived 
multi-frequency 
breather for $\omega = 2.7$ and $N = 41$ (black: -1, white: 1).}
  \label{mperdyn}
  \end{figure}

To confirm the stability results we turn to simulations of the dynamics.  In Fig. \ref{mperdyn}(a) we see the effect of the weak instability on the on-site out-of-phase breather for $N=101$.  After an extended period with little change the breather suddenly evaporates, with the system converging to the in-phase period-1 solution for all pendulums.  In contrast, at $\omega = 2.7$ for the breather in a chain with $N=41$ we see persistence over long simulation times (see Fig. \ref{mperdyn}(b)) suggesting the state is stable, in agreement with the stability results shown in Fig. \ref{mperfamN41}.
   
\subsection{Breather generation}
\label{generation}

\begin{figure}[htbp]
\centering
 \begin{minipage}{0.95\columnwidth}
  \centering
  \includegraphics[width=0.95\columnwidth]{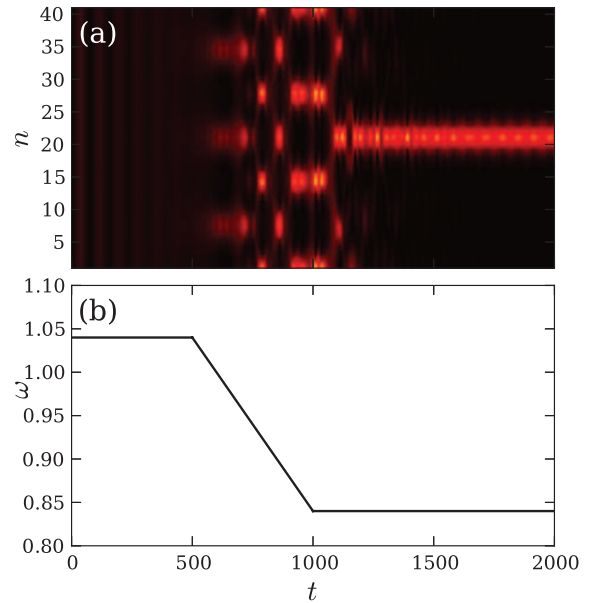}
  \end{minipage}
  \caption{(color online).  On-site breather generation from an initially quiescent pendulum configuration ($\theta(0) = \dot{\theta}(0) = 0$), through modulation of the driving frequency. (a) Generation of single on-site breather following period of instability dynamics; (b) time dependence of driving frequency $\omega$.}
  \label{breathgen}
  \end{figure}

\begin{figure}[htbp]
\centering
 \begin{minipage}{0.95\columnwidth}
  \centering
  \includegraphics[width=0.95\columnwidth]{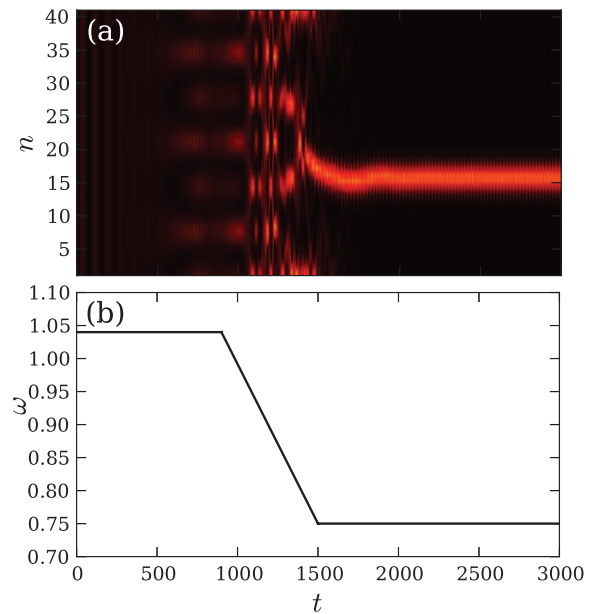}
  \end{minipage}
  \caption{(color online).  Off-site breather generation from an initially quiescent pendulum configuration ($\theta(0) = \dot{\theta}(0) = 0$), through modulation of the driving frequency. (a) Generation of single off-site breather following a period of unstable dynamics; (b) time dependence of driving frequency $\omega$.}
  \label{breathgenoff}
  \end{figure}

An important issue concerning breathers is how to generate them in practice.  We consider one possible scheme involving modulation of the driving frequency.  We propose starting with all pendulums at rest and driving at a frequency at which modulational instability emerges with a well-defined instability wavenumber.  In this way a regular pattern emerges spontaneously in the system.  We then suggest lowering the frequency until it sits within the stable region of the target breather type.  In Fig. \ref{breathgen} we show generation of a stable on-site breather, through variation of $\omega$ from $\omega = 1.04$ down to $\omega = 0.84$ (as depicted in Fig. \ref{breathgen}(b)).  If we lower the frequency further we sit in the band of stable off-site breathers, and we see in Fig. \ref{breathgenoff} that similarly we can generate members of this family of breathers.  In this latter case, we lower the frequency more rapidly from $\omega = 1.04$ to $\omega = 0.75$.  As expected, when relying on instability dynamics, we find the final configuration depends on the initial period at the unstable frequency, and the speed of the switch.  However, we find that the generation of breathers through this process is robust and occurs without any special choice of these parameters.  We should note that we have been unsuccessful in using this method to generate mixed-frequency breathers.  This appears to be due to the relatively small windows of stability (and presumably also corresponding basins of
attraction) of the breathers for our choice of parameters.  To achieve 
generation we have instead used initial conditions near the breather state.  
The general problem of the nature of the dynamics for different initial 
conditions is largely still an open one.

\section{Conclusions}

We have examined in detail the response of both a single pendulum and a
pendulum chain to horizontal driving of the pendulum pivot point, in an experimentally accessible region of the system parameter space, focusing on small driving amplitude.  We have characterized the single pendulum solutions, finding a region of bistability near the linear resonance of the pendulum natural frequency and driving frequency ($\omega = 1$), with one stable solution of large amplitude and roughly in phase with the driving force and the second stable solution of small amplitude and out of phase.  We find also the existence of solutions three times the period of the driving force, near the subharmonic resonance at $\omega = 3$.  These are large amplitude solutions, with no connection to the period-1 solutions.  Turning to the pendulum chain we show that the period-1 higher amplitude solution exhibits a modulational instability below a certain critical frequency, and beyond a certain chain length.  We characterize this modulational instability by studying the instability of a sinusoidal solution ansatz, for a low order expansion of the sinusoidal nonlinearity, and find good agreement with our numerical results.  We examine the on-site and off-site localized breather solutions in the pendulum chain, since the breather
waveforms appear to spontaneously form as a result
of this instability. More specifically, we show their close connection to the single pendulum periodic solutions.  In particular, the breathers only exist for parameters for which there is single pendulum bistability.  Following this we examine the possibility of multi-frequency breathers, corresponding to the bistable region near the first subharmonic resonance.  We find that for moderate chain lengths stable breathers exist in which the central sites are moving at one third of the frequency of the edge sites.  For longer chain lengths however we find these breathers are weakly unstable.  Finally we turn to the possibility of breather generation through dynamically tuning the driving frequency.  We discuss a robust scheme which may be used to generate both on-site and off-site breathers.  
Natural directions for future work include further characterization of the subharmonic response, possibly in connection with shorter chain settings and
the potential connection of this setting with experiments such as the
mechanical ones of~\cite{CuevasPRL2009}, as well as the electrical ones
of~\cite{lars2}. On the other hand, understanding more systematically
breather, as well as multi-breather states and a potential tuning
of their existence, as well as stability regimes would be of particular
interest in inducing  (and optimizing) energy localization in this
system. 



\end{document}